\documentclass[preprint,aps,prd,nofootinbib,floatfix,amsmath,amssymb]{revtex4}
\usepackage{graphicx}
\usepackage[utf8x]{inputenc}
\usepackage{color}
\usepackage[tight,TABTOPCAP]{subfigure}
\usepackage{multirow}
\usepackage{slashed}
\usepackage{cancel}%\cancel{} \bcancel{} \xcancel{} \cancelto{value}{expression}
\usepackage{hyperref}
\usepackage{amsmath}
\begin{document}
\markboth{P. L. C\'astulo,  
	J. J. Toscano, E. S. Tututi}{Effects of Lorentz violation in the Higgs sector of the Minimal  Standard Model Extension}

%\catchline{}{}{}{}{}
%
%%%%%%%%%%%%%%%%%%%%%%%%%%%%%%%%%%%%%%%%%%%%%%%%%%%%%%%%%%%%%%%%%%%%

\makeatletter
%Feynman slash
\newbox\slashbox \setbox\slashbox=\hbox{$/$}
\newbox\Slashbox \setbox\Slashbox=\hbox{\large$/$}
\def\pFMslash#1{\setbox\@tempboxa=\hbox{$#1$}
  \@tempdima=0.5\wd\slashbox \advance\@tempdima 0.5\wd\@tempboxa
  \copy\slashbox \kern-\@tempdima \box\@tempboxa}
\def\pFMSlash#1{\setbox\@tempboxa=\hbox{$#1$}
  \@tempdima=0.5\wd\Slashbox \advance\@tempdima 0.5\wd\@tempboxa
  \copy\Slashbox \kern-\@tempdima \box\@tempboxa}
\def\FMslash{\protect\pFMslash}
\def\FMSlash{\protect\pFMSlash}
\def\miss#1{\ifmmode{/\mkern-11mu #1}\else{${/\mkern-11mu #1}$}\fi}
%%%% Uso:  \pFMSlash{p}
\makeatother

%\tightenlines
\title{Effects of Lorentz violation in the Higgs sector of the Minimal  Standard Model Extension}
\author{P. L. C\'astulo$^{(a)}$,  
J. J. Toscano$^{(b)}$, E. S. Tututi$^{(a)}$}
\address{
$^{(a)}$Facultad de Ciencias F\'\i sico Matem\' aticas,
Universidad Michoacana de San Nicol\' as de
Hidalgo, Avenida Francisco J. M\' ujica S/N, 58060, Morelia, Michoac\'an, M\' exico. \\
$^{(b)}$Facultad de Ciencias F\'{\i}sico Matem\'aticas,
Benem\'erita Universidad Aut\'onoma de Puebla, Apartado Postal
1152, Puebla, Puebla, M\'exico.\\
pcastulo@gmail.com, jesus.toscano@correo.buap.mx, eduardo.tututi@umich.mx}

%\begin{history}
%	\received{Day Month Year}
%	\revised{Day Month Year}
%\end{history}

\begin{abstract}
A bound on the  CPT-odd  four vector coefficient $k^\mu_\phi$   that appears in   Higgs sector of the  minimal Standard Model Extended (mSME) is presented. The analysis  is based on the contributions arising from the sector in question  to the anomalous magnetic dipole moment  (AMDM) for leptons calculated at the one loop level, for which an analytical expression is obtained. The largest contribution of this Lorentz violating coefficient is on the lightest lepton, which results as a consequence of a strong non-decoupling effect. By using the experimental uncertainty of the electron AMDM we predict that $|k^2_{ \phi\,R}|<3.29\times 10^{-29}$ GeV$^2$. 

\end{abstract}

\maketitle

%%\pacs{12.60.Cn, 14.70.Fm, 13.40.Em, 11.30.E}

\section{Introduction}\label{i} 

Currently, there are well established reasons to believe that the  Lorentz Symmetry could not be satisfied at very high energies or equivalently at short distances. This has raised the interest in  searching for signals to confirm whether there exists Lorentz Symmetry violation. The Standard Model Extension (SME) is an effective field theory formulation that encompasses the Standard Model (SM)  coupled to  general relativity along with all possible field operators for Lorentz violation~\cite{SME-1,SME-2}. This effective theory predicts, for instance,  unconventional physics effects such as birefringence \cite{SME}, anisotropies in particle dispersion relations, modified particle kinematics as well as modified particle dynamics, etc. \cite{colladay1,kostelecky}, or  it can give  alternative, although even perhaps possible, explanation to the neutrino oscillation phenomenon, based on Lorentz and CPT violation \cite{netri-oscil-cpt,netri-oscil-cpt-1,netri-oscil-cpt-2,netri-oscil-cpt-3,netri-oscil-cpt-4}.  However, currently no one experimental evidence exits that corroborates the Lorentz symmetry violation.  The full Lagrangian of the SME contains renormalizable and non-renormalizable Lorentz violating terms that  are  the result of products of field  operators  with coefficients independent of  coordinates in such a way that any experimental signal for Lorentz violation, can be expressed in terms of one or more of these coefficients. The field operators are  classified according to their  mass dimension.  A restricted case is the minimal Standard Model Extension (mSME) which is constructed with field operators of  mass dimension 4 or less \cite{SME}.  
The SME contains  an infinity quantity of Lorentz-violating coefficients in: the  matter, gauge, Higgs and gravity sectors. A lot of these coefficients have been studied by means of different terrestrial high precision experiments or astrophysical observations. Concerning the mSME, many of the aforementioned coefficients   are reported in Ref. \cite{sme-rmf} which is continuously updated.

There are other sectors of the mSME, such as the Higgs sector in which some of its coefficients  have also been studied in various scenarios of high energy physics  at the tree \cite{aranda1,aranda2} and one loop level \cite{anderson,toscano1,toscano2,toscano3}. In Ref.	\cite{anderson} it is studied the Higgs sector, where it was established bounds on the    CPT-even and CPT-odd coefficients appearing in this sector, by considering the photon and $Z$ gauge boson propagators, respectively. The  $k_{\phi \phi}^{\mu\nu}$ CPT-even anti-symmetric coefficients, were broadly discussed, also   bounds for these were  established  in \cite{anderson,toscano1}. In particular, in \cite{anderson} the CPT-odd coefficient  $k_\phi^\mu$ was bounded in an indirect way, by considering a non zero expectation value of the $Z_\mu$ gauge boson field.  
The best bound obtained for  $\mathrm{Re}(k_\phi^\mu)$ is less than $10^{-31}$ GeV. This bound is derived from neutrons with the use of two-species noble-gas maser \cite{maser}.
In Ref. \cite{toscano1}, it was   shown that $\mathrm{Im}\,k_{\phi \phi}^{\mu\nu}$,  can be removed out in favor of others terms in the same CPT-even Higgs sector of the mSME.

In this paper we also analyze the CPT-odd coefficient in Higgs sector of the mSME through the  $\bar{l}l\gamma$ vertex at the one-loop level, where $l=e,\mu,\tau$. In particular, we focus on  finding bounds for the scalar product of $k_\phi^\mu$ four-vector  in the Higgs sector of the mSME. This is done   by calculating the contribution to the dipolar magnetic moment of the lepton according to sector in question.  Since  $k_\phi^\mu$ has mass dimension, it results an non-decoupling effect on physical quantities, such as the dipolar  magnetic moment. The result of this  effect is that the amplitude behaves as $1/m_e^4$. We use this result along with the experimental measurements of high precision of the AMDM of the electron to do numerical predictions.

An outline of this paper is as follows. In Sec. II we present the piece of  Lagrangian to be used in the work. In Sec. III we develop the calculations of the CPT-odd dipolar magnetic moment of leptons. Finally, in Sec IV we give our final considerations.

\section{Theoretical framework}

 \subsection{General structure of the $\Gamma^\mu$ vertex }

 In order to carry out the calculation of  $\bar{l}l\gamma$ coupling at the one loop level, let us consider the most  general Lorentz structure of the vertex in question that couples a photon with charged leptons on shell in terms of independent form factors.  The vertex  function can be expressed as follows  \cite{nowakowski}:
 \begin{equation}
 	\Gamma_\mu(q^2)= F_1(q^2)\gamma_\mu+ F_2(q^2)i\sigma_{\mu\nu} q^\nu -F_3(q^2)\gamma_5\sigma_{\mu\nu}q^\nu ,
 	\label{general-vertex}
 \end{equation}
 where, as usual, $q=p-p^\prime$ is the  transferred momentum  and $\sigma^{\mu\nu}\equiv \frac{i}{2}[\gamma^\mu, \gamma^\nu]$. As it is well known, the  $F_1(0)=Q_l$ form factor is related to the electric charge of the lepton, the  $F_2$ and $F_3$ form factors are related to the  dipole moments of the lepton $l$ with mass $m_l$ as follows:
 \begin{align}\label{Dipole-moments}
 	a_l= -2 m_l F_2(0), \qquad      d_l=  F_3(0),
 \end{align}
where, as usual, $a_l$ and $d_l$ stand for the AMDM and  the electric dipole moment of the lepton $l$, respectively. The expressions in Eq. \ref{Dipole-moments}   will be used in what follows.

\subsection{Higgs sector of the SME}\label{L}
In order to simplify our analysis as much as possible, we restrict ourselves to the Higgs sector of the mSME \cite{SME}. The CPT violation in this sector occurs  through the following term 
\begin{equation}
{\cal L}_{\mathrm{Higgs}}^\mathrm{CPT-odd}=ik_\phi^\mu\left(\phi^\dagger D_\mu \phi\right) +\mathrm{H.c.},
\label{leffec1}
\end{equation}
where $\phi$ is the $SU(2)$-doublet  Higgs field, $k_\phi^\mu=k_{\phi \,R}^\mu+ik_{\phi\, I}^\mu$ is constant complex  four-vector with dimensions of energy, being $k_{\phi \,R,I}^\mu$ its real and imaginary parts respectively. The covariant derivative $D_\mu$ is 
\begin{equation}
D_\mu=\partial_\mu -i\frac{g}{\sqrt{2}}\left(\sigma^+W^+_\mu+\sigma^-W^-_\mu\right)-i\frac{g}{2}\cos\theta_W\left(\sigma^3-\tan^2\theta_W Y\right)Z_\mu-ieQA_\mu,
\end{equation}
where $g$ is the weak constant coupling, $\sigma^{\pm}=1/2(\sigma^1\pm i\sigma^2)$, $\theta_W$ is the weak angle, $Y$ the hypercharge, and  $Q=1/2(\sigma^3+ Y)$, being $\sigma^i, i=1,2,3$ the Pauli matrices.   In the unitary gauge, we can work out the Lagrangian in Eq. (\ref*{leffec1}) to obtain
\begin{equation}
{\cal L}_{\mathrm{Higgs}}^\mathrm{CPT-odd}=ik_{\phi I}^\mu\left(v+H\right)\partial_\mu H-k_{\phi R}^\mu\frac{g}{2\cos\theta_W}\left(v+H\right)^2Z_\mu.
\label{leffec2}
\end{equation}
Notice that we can  drop the term containing the factor $H\partial_\mu H=1/2\partial_\mu(H^2)$ in last equation, since it represents a surface term. The Lagrangian in (\ref{leffec2}) will be useful for our purposes, from which we can extract the Feynman rules that will be used in the calculation presented below. In particular, let us show the
Feynman rule that, in addition to the used in the electroweak sector  of the SM, will be employed in the calculation.  This corresponds to a line of a Higgs connecting a line of a $Z$ gauge boson as follows:
\begin{figure}[htb!]
\centering
\includegraphics[scale=0.65]{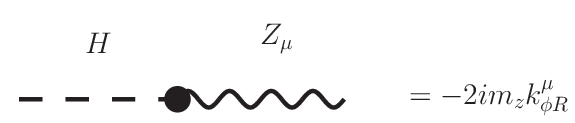}
\caption{Feynman rule.}
\label{diagram0}
\end{figure}

\section{The CPT-odd anomalous magnetic dipole moment} 
\begin{figure}[htb!]
\centering
\subfigure[]{\includegraphics[scale=0.65]{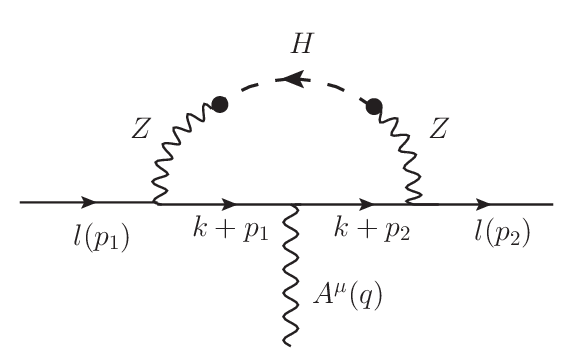}}\qquad
\subfigure[]{\includegraphics[scale=0.65]{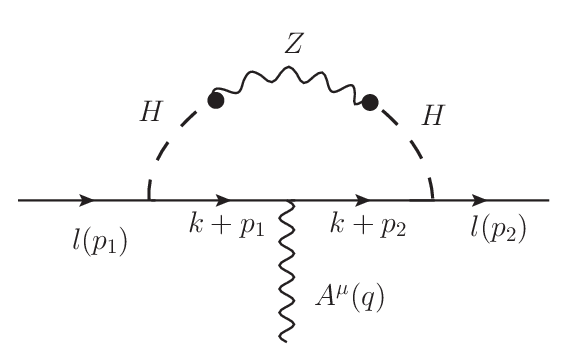}}
\caption{Feynman diagrams contributing to the  $\bar{l}l\gamma$ vertex.}
\label{diagrams}
\end{figure}

Having presented the structure of the Lagrangian of interest,  we now turn to present the calculation of the vertex  that involves the information on the $k_{\phi\,R}^\mu$ four vector.	
Let us consider the invariant amplitude 

\begin{equation}
-i{\cal M}=\bar{u}(p_2)\left(-i\Gamma_\mu\right) u (p_1)\epsilon^{\mu *}(q),
\label{amplitude}
\end{equation}
where the vertex $\Gamma_\mu=\Gamma_\mu^{(a)}+\Gamma_\mu^{(b)}$ receives contributions at the one loop level  from diagrams  in Fig. \ref{diagrams} (a) and (b). For the $\Gamma_\mu^{(a)}$ contribution, we have
\begin{equation}
-i\Gamma_\mu^{(a)}=i\frac{eg^2m_Z^2}{\cos^2\theta_W}\int\frac{d^4k}{(2\pi)^4}\frac{T_\mu}{\left[k^2-m^2_H\right]\left[k^2-m^2_Z\right]^2\left[\left(k+p_1\right)^2-m^2_l\right]\left[\left(k+p_2\right)^2-m^2_l\right]},
\label{vertexa}
\end{equation}
with
\begin{eqnarray}
T_\mu & = &(k_{\phi R})_\lambda(k_{\phi R})_\beta\left(g^{\alpha\beta}-
\frac{k^\alpha k^\beta}{m_Z^2}\right)\left(g^{\beta\lambda}-
\frac{k^\beta k^\lambda}{m_Z^2}\right)
\bigg[\gamma_\beta\left(g_V^f-g_A^f\gamma^5\right)
\nonumber \\
\ && \times (\pFMSlash{k}+\pFMSlash{p}_2+m_l)\gamma_\mu (\pFMSlash{k}+\pFMSlash{p}_1+m_l)\gamma_\alpha \left(g_V^f-g_A^f\gamma^5\right)\bigg].
\label{nume}
\end{eqnarray}
The contribution resulting from diagram  in Fig. \ref{diagrams} (b) is quite suppressed  due to the coupling $H\bar{l}l$ which is $\frac{igm_l}{2m_W}$, hence the corresponding amplitude is suppressed by a factor $\left(\frac{m_l}{m_W}\right)^2$ and, consequently, we can neglect it. 

In order to carry out the calculations of the invariant amplitude ${\cal M}$ in Eq. (\ref{amplitude}) and the vertex function, $\Gamma^\mu$, we use the tensor decomposition of \texttt{FeynCalc} \cite{feyncalc} to express the form factors in Eq. (\ref{general-vertex}) in terms of Passarino-Veltman (PV) scalar functions. The result is quite lengthy and it depends on  $A_0$,  $B_0$, $C_0$ and $D_0$ PV scalar functions, in general with different arguments.   Let us comment that at some stage of calculation, we  used different identities of PV scalar functions such as   $A_0(m^2)=m^2[1+B_0(0,m^2,m^2)]$ in order to eliminate the $A_0$, $C_0$ and  $D_0$  functions with proper arguments in favor of the $B_0$'s functions.
Finally, the contribution due to CPT-odd coefficients to the AMDM, $a_l^{\mathrm{CPT-odd}}$ for a lepton $l$, can be cast as:
\begin{equation}
a_l^{\mathrm{CPT-odd}}=\frac{\alpha}{\pi\sin 2\theta_W}\left(\frac{m_Z^2}{m_H^2-m_Z^2}\right)\left(\frac{k_{\phi_R}^2}{m_l^2}\right)\left[(g_V^l)^2f_V(0)+(g_A^l)^2f_A(0)\right],
\label{dipole}
\end{equation}
where, as usual, $\alpha=e^2/4\pi$ is the fine structure constant.   We recall that $k_{\phi_R}^2=k_{\phi_R}^\mu k_{{\phi_R}\, \mu}$ and it can be positive, negative or  zero. In the last case, nothing to do. Thus, we tacitly assume that   $k_{\phi_R}^\mu$ can be either space-like or time-like four-vector. The vector  $f_{V}(0)=f_{V}(q^2=0)$ and the  axial  $f_A(0)=f_A(q^2=0)$  form factors are given respectively by
\begin{equation}
f_{V,A}(0)=f_0^{V,A}+\sum_{i=1}^5f_i^{V,A}B_0(i)+f_6^{V,A}m_Z^2C_0.
\label{form-factors}
\end{equation}
Here,  the various $B_0(i)$ functions stand for PV's with different argument. The explicit expressions for the  functions involved in this equation are displayed  in the appendix. Although  the dipole magnetic moment in Eq. (\ref{dipole}) is expressed in a linear form, as a function of  different $B_0$'s, each containing  an  UV divergence,  its is easy to see that the final result is UV free, as it must be.

An analysis of the different functions in Eq. (\ref{form-factors}), indicates that the $a_l^{\mathrm{CPT-odd}}$ is very sensitive to the mass $m_l$ of the charged lepton. This implies that the main result emerges from the AMDM of the electron. The most important contributions to the vector and axial parts come from the terms $f_1^{V,A}B_0(1)$, $f_2^{V,A}B_0(2)$ and $f_6^{V,A}m_Z^2C_0$, respectively. Moreover a Taylor expansion around $m_l=0$  shows that the dominant contribution to $f_V(0)$ part is
\begin{equation}
f_V(0)=\frac{3}{8}\left(\frac{m_Z^2}{m_e^2}\right)\log\left(\frac{m_Z^2}{m_e^2}\right)+\cdots,
\label{vector-fm-red}
\end{equation}
where the ellipsis stand for contributions ${\cal O}(m_e^2/m_Z^2)$ which can be dropped. Analogously,  the dominant contribution  to the axial form factor comes  from the  electron:
\begin{equation}
f_A(0)=\frac{9}{32}\left(\frac{m_Z^2}{m_e^2}\right)\log\left(\frac{m_Z^2}{m_e^2}\right)+\cdots.
\label{axial-fm-red}
\end{equation}
Hence, the main contribution to the electron AMDM arising from CPT-odd Higgs sector of the mSME  can be written as
\begin{equation}
a_e^{\mathrm{CTP-odd}}=\frac{\alpha}{\pi\sin^22\theta_W}\left(\frac{m_Z^2}{m_H^2-m_Z^2}\right)\left(\frac{|k_{\phi_R}^2|}{m_e^2}\right)\left[\frac{3}{8}\left(g_V^e\right)^2+\frac{9}{32}\left(g_A^e\right)^2\right]\left(\frac{m_Z^2}{m_e^2}\right)\log\left(\frac{m_Z^2}{m_e^2}\right),
\label{final-result}
\end{equation}
where, for convenience and  to compare with the experimental value of the AMDM of the electron, we have taken the absolute value of $k_{\phi, R}^2$. By demanding that $a_e^{\mathrm{CTP-odd}}$  be less than the experimental uncertainty:  $a_e^{\mathrm{CTP-odd}}<\Delta a_e^{\mathrm{Exp}}$, we can establish a bound for the CPT-odd  coefficient: 
\begin{equation}
|k_{\phi_R}^2|<\left\{\frac{\alpha}{\pi\sin^22\theta_W}\left(\frac{m_Z^4}{m_H^2-m_Z^2}\right)\left[\frac{3}{8}\left(g_V^e\right)^2+\frac{9}{32}\left(g_A^e\right)^2\right]\log\left(\frac{m_Z^2}{m_e^2}\right)\right\}^{-1}m_e^4\,\,\Delta a_e^{\mathrm{Exp}}.
\label{final-bound}
\end{equation}
In order to determine the numerical bound, we use the values reported in the Particle Data Group  \cite{pdg} for different parameter in the expression, finding that $|k_{\phi_R}^2|<3.29\times 10^{-29}$ GeV$^2$.

\section{Conclusions and final considerations}
The mSME is a minimal extension to the standard model that accounts for the violations of the Lorentz  and CTP  symmetries that could occur in nature. The corresponding Lagrangian contains terms composed by CPT-even and CPT-odd coefficients that couple to matter and gauge fields. The knowledge of these coefficients  could be  helpful to find out in a more exhaustive way  the possible violations of the mentioned symmetries. In this paper we have presented a calculation on the contributions arising from the CPT-odd Higgs sector of the mSME to  the  AMDM of charged leptons. We found that, in the limit $m_l\ll m_H, m_Z$, the dominant contribution is due to the lighter lepton. For the case of an electron, we obtained an analytical expression for the AMDM  that takes into account the effects of  the CPT-odd background resulting from the sector under consideration.  This allowed us to establish the bound on the CPT-odd coefficient:  $|\,k^2_{\phi\,R}|<3.29\times 10^{-29}$ GeV$^2$.

\section*{Acknowledgments}
{We acknowledge financial support from CIC-UMSNH and
SNI (M\'exico).}

\appendix
\section{} 
Useful relation between Passarino-Veltman scalar functions.
\begin{align*}
B_0(1)=&B_0(m_l^2,m_l^2,m_Z^2)
\\
B_0(2)=&B_0(m_l^2,m_l^2,m_H^2)
\\
B_0(3)=&B_0(0,m_Z^2,m_Z^2)
\\
B_0(4)=&B_0(0,m_H^2,m_H^2)
\\
B_0(5)=&B_0(0,m_H^2,m_H^2)
\\
C_0 \equiv & C_0(m_l^2,0,0,m_Z^2,m_l^2,m_Z^2)
\\
=&\frac{\partial}{\partial m_Z^2}\frac{1}{i\pi^2}\int d^4k\frac{1}{(k^2-m_Z^2)[(k+p_1)^2-m_l^2]}
\\
=&\frac{1}{m_l^2\sqrt{1-\frac{4m_l^2}{m_Z^2}}}\left[-\frac{1}{2}\sqrt{1-\frac{4m_l^2}{m_Z^2}}\log\left(\frac{m_Z^2}{m_l^2}\right)+\left(1-\frac{2m_l^2}{m_Z^2}\right)\log\left(\frac{1-\sqrt{1+\frac{4m_l^2}{m_Z^2}}}{1-\sqrt{1-\frac{4m_l^2}{m_Z^2}}}\right)\right]
\end{align*}

Vector form factors:
\begin{align*}
f_0^V=&\frac{1}{(m_H^2-4m_l^2)(m_Z^2-4m_l^2)^2(m_H^2-m_Z^2)}\bigg\{-\frac{7}{4}m_H^2m_Z^4(m_H^2-m_Z^2)\\
&+m_l^2\left[4m_l^2(m_H^2-m_Z^2)(16m_l^2-15m_Z^2)+m_Z^2(11m_H^4-4m_Z^2m_H^2-7m_Z^4)\right]\bigg\}
\\
f_1^V=&\frac{1}{4(m_Z^2-4m_l^2)^2(m_H^2-m_Z^2)}\bigg\{m_Z^4(13m_H^2-5m_Z^2)\\
&+2m_l^2\left[4m_l^2(17m_H^2-5m_Z^2)-m_Z^2(47m_H^2-19m_Z^2)\right]\bigg\}
\\
f_2^V=&-\frac{2m_H^2(m_H^2-3m_l^2)}{(m_H^2-m_l^2)(m_H^2-m_Z^2)}
\\
f_3^V=&-\frac{1}{4(m_Z^2-4m_l^2)^2(m_H^2-m_Z^2)}\bigg\{m_Z^4(13m_H^2-5m_Z^2)\\
&+2m_l^2\left[m_l^2m_H^2+12m_Z^2(3m_Z^2-7m_H^2)\right]\bigg\}
\\
f_4^V=&\frac{m_l^2}{2(m_H^2-4m_l^2)(m_Z^2-4m_l^2)^2}\bigg\{4m_l^2(20m_l^2-9m_H^2-m_Z^2)
+5m_H^2m_Z^2\bigg\}
\\
f_5^V=&\frac{2m_H^2(m_H^2-2m_l^2)}{(m_H^2-4m_l^2)(m_H^2-m_Z^2)}
\\
f_6^V=&\frac{3m_Z^2-4m_l^2}{4(m_Z^2-4m_l^2)}
\end{align*}

Axial form factors:
\begin{align*}
f_0^A=&\frac{1}{48m_Z^4(m_Z^2-4m_l^2)}\bigg\{m_Z^2\left[16m_H^2(3m_H^2-m_Z^2)-85m_Z^4\right]\\
&+2m_l^2\left[4m_l^2(33m_H^2-73m_Z^2)-m_H^2(96m_H^2+m_Z^2)+195m_Z^4\right]\bigg\}
\\
f_1^A=&\frac{1}{48m_Z^2(m_Z^2-4m_l^2)^2(m_H^2-m_Z^2)}\bigg\{5m_Z^4(33m_H^2-17m_Z^2)\\
&+2m_l^2\left[4m_l^2(93m_H^2-29m_Z^2)-m_Z^2(399m_H^2-175m_Z^2)\right]\bigg\}
\\
f_2^A=&\frac{1}{48m_Z^4(m_H^2-m_Z^2)}\bigg\{-16m_H^2(3m_H^4-4m_H^2m_Z^2+6m_Z^4)\\
&+32m_l^2(3m_H^4-2m_H^2m_Z^2+3m_Z^4)\bigg\}
\\
f_3^A=&\frac{1}{48m_Z^2(m_Z^2-4m_l^2)(m_H^2-m_Z^2)}\bigg\{-5m_Z^4(33m_H^2-17m_Z^2)\\
&-4m_l^2\left[48m_l^2(2m_H^2-m_Z^2)+m_Z^2(73m_Z^2+165m_H^2)\right]\bigg\}
\\
f_4^A=&\frac{m_l^2}{48m_Z^4(m_H^2-m_Z^2)}\bigg\{8m_l^2(24m_H^2-53m_Z^2) -2m_Z^2(24m_H^2-77m_Z^2)\bigg\}
\\
f_5^A=&\frac{1}{48m_Z^4(m_H^2-m_Z^2)}\bigg\{16m_H^2(3m_H^4-4m_H^2m_Z^2+6m_Z^4) -48m_l^2m_Z^4\bigg\}
\\
f_6^A=&\frac{9}{16}-\frac{m_l^2}{4m_Z^2}
\end{align*}

\end{document}